\documentstyle{article}
\newcommand{\hochpunkt}[1]{\mbox{$^{\raisebox{.3ex}{\scriptsize #1}}_{\raisebox{.6ex}{\hspace{.17em}.}}$}}
\newcommand{\hoch}[1]{\mbox{$^{\raisebox{.3ex}{\scriptsize #1}}$}}
\begin{document}

\begin{center}
{\LARGE \bf Photometry of the long period dwarf nova 
MU~Centauri}\footnote{Based 
on observations taken at the Observat\'orio do Pico dos Dias / LNA}

\vspace{1cm}

{\Large \bf Albert Bruch}

\vspace{0.5cm}
Laborat\'orio Nacional de Astrof\'{i}sica, Rua Estados Unidos, 154,
CEP 37504-364, Itajub\'a - MG, Brazil
\vspace{1cm}

(Text published in: New Astronomy, Vol.\ 46, p.\ 60 -- 72 (2016))
\vspace{1cm}

{\bf Abstract}

\vspace{0.5cm}
\end{center}

\begin{quote}
{\parindent0em Even among the brigher cataclysmic variables an appreciable
number of objects exist about which not much is known. One of them, MU~Cen, 
was observed as part of a small project to better characterize these
neglected systems. The temporal variations of the brightness of MU~Cen 
during quiescence were studied in order to find clues to the structure
of the system and its behaviour on time scales of hours and shorter.
Light curves observed in white light at a time resolution of a 
few seconds and with a duration of several hours, obtained in six nights
and spanning a total time base of five months, were investigated using 
different time series analysis tools, as well as model fits.
The light curve of MU~Cen is dominated by ellipsoidal variations of the 
secondary star. The refined orbital period is $P_{\rm orb} = 0.341883$~days.
Model fits permit to constrain the temperature of the secondary 
star to $\sim$5000~K and the orbital inclination to $50^o \le i \le 65^o$. The
latter result permits estimates of the component masses which are probably 
somewhat smaller that derived in previous publications.
A second persistent period of $P_2 = 0.178692$~days was also
identified. Its origin remains unclear. As all cataclysmic variables, MU~Cen 
exhibits flickering, however, on a rather low level. Its frequency behaviour
is normal for quiescent dwarf novae. There are indications that the 
individual flickering events are not always independent but can lead to
effects reminiscent of quasi-periodic oscillations.}
\vspace{0.5cm}

{\parindent0em {\bf Keywords:}
Stars: binaries: close -- Stars: dwarf novae -- 
Stars: individual: MU~Cen}

\end{quote}

\section{Introduction}
\label{Introduction}

Cataclysmic variables (CVs) are interactive binaries where a late type, 
low mass star which is normally on or close to the main sequence transfers
matter to a white dwarf. Unless strong magnetic fields are present the
transferred matter first forms an accretion disk before it settles onto
the surface of the compact object. All CVs are expected to suffer from
major outbursts on secular time scales when they appear as novae. Depending
on the detailed conditions, many of them also undergo intermittent
less violent outbursts on time scale of weeks, months or years. These are the
dwarf novae. 

The number of known systems of this kind has grown enormously in recent
years. Much of this growth is due to many CVs detected in
large scale surveys either as a by-product of their main purpose (e.g., the
Sloan Digital Sky Survey) or as results of dedicated searches for transient
sources (e.g. Catalina Real-time Transient Survey; MASTER Global Robotic Net). 
Most of these newly detected
CVs are rather faint in their normal brightness state which, in the case of
dwarf novae, is the quiescent phase between outbursts. Therefore, the
characterization of their individual properties is expensive because is 
requires large telescopes.

On the other hand, it is much easier to perform detailed studies of the
brighter CVs, most of which are known for a long time. It may therefore be
expected that these systems are all quite well known. However, this is not
the case.  Surprisingly, even after decades of intense observations
of CVs there is still an appreciable number of known or suspected
systems, bright enough (say, $m_{\rm vis} \le 15\hoch{m}$)
to be easily observed with comparatively small
telescopes, which have not been studied sufficiently to even be certain
-- in some cases -- of their very class membership. I therefore started
a small observing program aimed at a better understanding of these so far
neglected stars. Here, I report on the results concerning the dwarf
nova MU~Cen.

\begin{figure}
   \parbox[]{0.1cm}{\epsfxsize=14cm\epsfbox{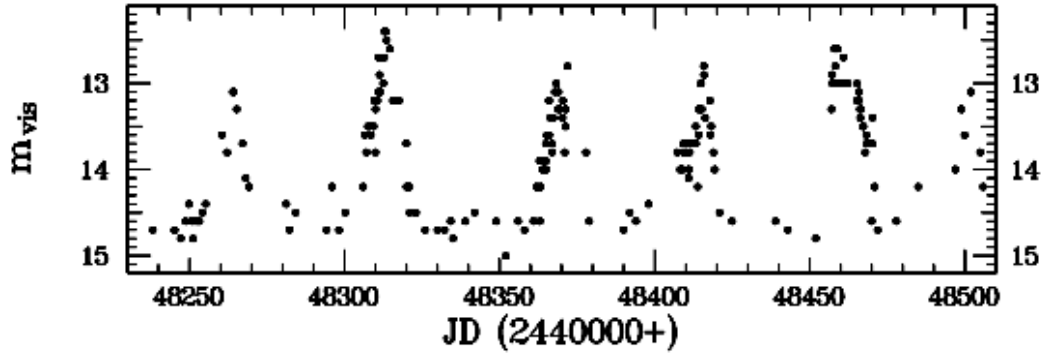}}
      \caption[]{Long term light curve
                 of MU~Cen as observed by AAVSO members during the
                 1991 observing season.}
\label{mucen-aavso}
    \end{figure}

In this case there is no doubt about the class membership.
MU~Cen is well established as a dwarf nova. Fig.~\ref{mucen-aavso} 
shows a section of the long-term AAVSO light curve which encompasses
the entire 1991 observing season. The typical alternations between quiescent
states at a level of $\sim$$14\hochpunkt{m}6$ and outbursts reaching
$\sim$$12\hochpunkt{m}4$ leave no doubts about the classification. Colours in 
the $UBV$ system measured by Mumford (1971) and 
Vogt (1983) are comparatively red for a cataclysmic variable 
(see Bruch \& Engel, 1994), indicating a significant contribution
of the cool secondary star and thus a long orbital period. 

\input epsf

Descriptions of the spectral characteristics
of MU~Cen are somewhat ambiguous. Vogt (1976) describes the
spectrum at minimum stage as a continuum without strong emission features.
Citing a private communication from M.W.\ Feast, Warner (1976) describes
it to be weak, containing narrow hydrogen and possibly He~II lines in
emission, again during minimum. Vogt's description contrast somewhat 
with the optical spectrum reproduced in Fig.~14 of Zwitter \& Munari (1996) 
which is fairly normal for a quiescent dwarf nova and is in agreement
with  Warner's characterization, except that it does not show any trace of 
He~II. 

On the basis of a relation between the colours and orbital periods of
quiescent dwarf novae and novae, Vogt (1981) first estimated the 
period of MU~Cen to be $8.8$ hours. Later, Friend et al.\ (1990), 
using radial 
velocity measurements of absorption features of the secondary star observed 
in near infrared spectra in three successive nights, derived a value of 
$P_{\rm spec} = 0.342 \pm 0.001$ 
days ($8.208 \pm 0.024$ hours)\footnote{I denote the spectroscopic period
of Friend et al.\ (1990) by $P_{\rm spec}$ in order to distinguish it 
from the refined orbital period $P_{\rm orb}$ defined later.}. 

Time resolved photometric observations of MU~Cen are scarce. In particular, 
no light curve with a time resolution sufficient to show 
flickering has ever been published. 
It is well known that accretion of mass via a disk onto a central object
normally leads to apparently stochastic brightness variations termed
flickering. In CVs they cause variability on timescales typically of the order 
of minutes and with amplitudes which can range from a few millimagnitudes to 
more than an entire magnitude. For a general characterization of flickering 
in CVs, see Bruch (1992). The only photometric measurements of MU~Cen
consisting of more than isolated brightness measurements have been performed 
by Echevarr\'{\i}a (1988) at a time resolution of $\sim$4-5 minutes
during the rise to an outburst on 1986, March 8. Disregarding the long term
brightness increase over the $\sim$3 hours of observations, the scatter of
the data points in the $V$ band remain within $\sim$$0\hochpunkt{m}05$. In 
view of the coarse time resolution and the small amplitude it is not clear
if this scatter
is due to flickering. Therefore, the present observations of MU~Cen were
initiated with the objective to verify the presence of flickering in this
system. But, as will be seen, they revealed much more details about MU~Cen 
than just rapid and erratic brightness variations.

\section{Observations}
\label{Observations}

All observations were carried out at the 0.6-m Zeiss and the
0.6-m Boller \& Chivens telescopes of the Observat\'orio do Pico dos Dias, 
operated by the Laborat\'orio Nacional de Astrof\'{\i}sica, Brazil. In 
order to maximize the count rates in the short exposure intervals necessary 
to resolve the expected rapid flickering variations no filters were used.

Time series imaging of the field around MU~Cen was performed using cameras 
of type Andor iKon-L936-BV equipped with back illuminated, visually optimized
CCDs. Basic data reduction (biasing, flat-fielding) was performed
using IRAF. For the construction of light curves aperture photometry routines 
implemented in the MIRA software system (Bruch, 1993) were employed. The
same system was used for all further data reductions and calculations.
No attempt was made to calibrate the white light measurements. Brightness
was rather measured as magnitude difference of the target star with 
respect to several comparison stars in the field.
A summary of the observations is given in Table~\ref{Journal of observations}. 

\begin{table}
\caption{Journal of observations}
\label{Journal of observations}

\hspace{1ex}

\begin{tabular}{lcccc}
\hline
Obs. & Start & End    & Time  & Number  \\
Date & (UT)  & (UT)   & Res.  & of      \\
     &       &        & (s)   & Integr. \\
\hline
2015 Fev 11 & \phantom{2}6:00 & \phantom{2}7:58
          & \phantom{1}5\phantom{.5}  &           1\,394 \\ 
2015 May 21/33 &        21:35 & \phantom{2}3:15
          & \phantom{1}6\phantom{.5}  &           2\,728 \\ 
2015 Jun 08/09 &        21:59 & \phantom{2}2:17
          & \phantom{1}5\phantom{.5}  &           2\,989 \\ 
2015 Jun 09/10 &        21:26 & \phantom{2}1:42
          & \phantom{1}5\phantom{.5}  &           2\,679 \\ 
2015 Jun 10/11 &        21:17 & \phantom{2}1:13
          & \phantom{1}5\phantom{.5}  &           2\,761 \\ 
2015 Jun 11/12 &        21:18 & \phantom{2}1:38
          & \phantom{1}5\phantom{.5}  &           2\,988 \\ [1ex] 
\hline
\end{tabular}
\end{table}

\section{The light curve}
\label{The light curve}

\subsection{General properties}
\label{General properties}

I observed light curves of MU~Cen in six nights in 2015 (see
Table~1) at a time resolution of 5\hoch{s} -- 6\hoch{s}. 
UCAC4 228-062720 (taken from the Forth USNO CCD Astrograph Catalogue; 
Zacharias et al., 2013) was used as the primary comparison star. 
MU~Cen did not show large night-to-night variations. Since four of the
observing nights are contiguous this shows that the system was not in outburst.
A rough estimate, involving the magnitude of UCAC4 228-062720 quoted by
Zacharias et al.\ (2013) ($V=13\hochpunkt{m}031$ and 
$B=14\hochpunkt{m}022$; meaning that its spectral type is roughly K0),
the isophotal wavelength of the observations ($\sim$5960~{\AA}; right 
between the $V$ and $R$ bands; see Sect.~\ref{MU Cen Discussion}), 
the expected $V-R$ colour of the dominating MU~Cen secondary star 
(Sect.~\ref{MU Cen Discussion}), and the magnitude difference between
MU~Cen and UCAC4 228-062720, leads to a visual magnitude of MU~Cen
which is similar to the average quiescent visual magnitude (see 
Fig.~\ref{mucen-aavso}).

   \begin{figure}
   \parbox[]{0.1cm}{\epsfxsize=14cm\epsfbox{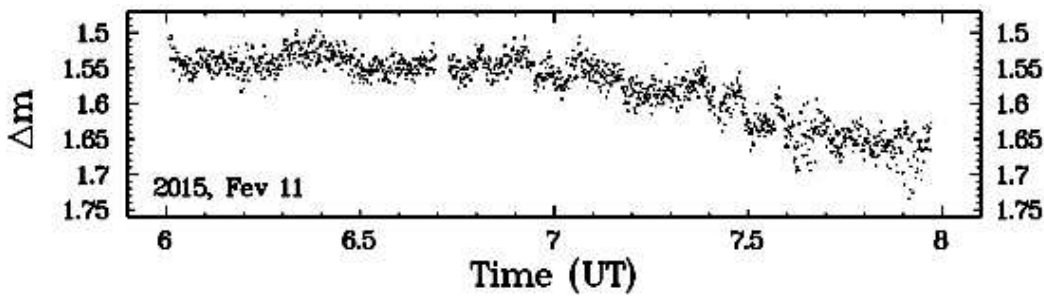}}
      \caption[]{Light curve of MU~Cen during the night of 2015, Feb.\ 11.}
\label{mucen-lc-1}
\end{figure}

\begin{figure}
   \parbox[]{0.1cm}{\epsfxsize=14cm\epsfbox{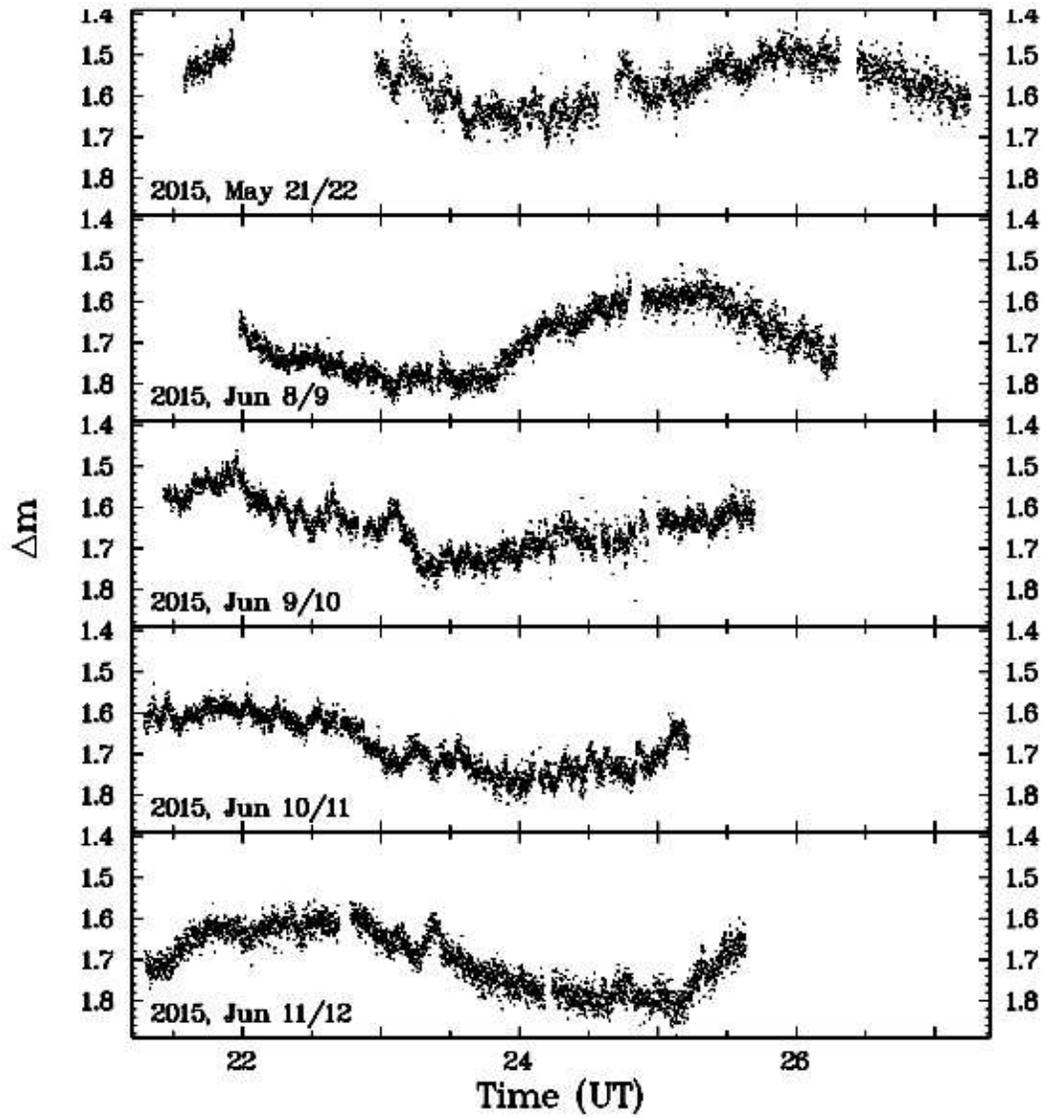}}
      \caption[]{Light curves of MU~Cen during several nights in 2015 May and
                 June, plotted on the same time and magnitude scale.} 
\label{mucen-lc-2}
\end{figure}

The (shorter) light curve of 2015, Feb.\ 11 is shown in Fig.~\ref{mucen-lc-1}.
Fig.~\ref{mucen-lc-2} contains the other light curves.
All of them show a clear modulation on a time
scale of about 4 hours, i.e., roughly half the spectroscopic period. This
suggest immediately ellipsoidal variations of the secondary star which 
should contribute a significant part of the light in this long period dwarf
nova. A more detailed analysis of the variations confirms this suspicion
(see below). Superposed on the modulation on hourly time scales are erratic
low amplitude short term fluctuations which are compatible with flickering on
a rather low scale. 

\subsection{Periods in the light curve}
\label{Periods in the light curve}

In order to characterize the variations on time scales of several hours
the light curves were combined into a single data set after appĺying the
baricentric correction. In order to
take into account possible night-to-night variations of the average 
brightness, I first performed a sine-fit to the original light 
curves\footnote{For the nights of February 11 and June 9 it was necessary to
fix the period to $P = 0.5 \times P_{\rm spec}$ to get consistent results;
during the other nights the sine-fit converged to a period very close 
to $0.5 \times P_{\rm spec}$.} and subtracted the vertical offset of the best
fit sine curve. 

   \begin{figure}
   \parbox[]{0.1cm}{\epsfxsize=14cm\epsfbox{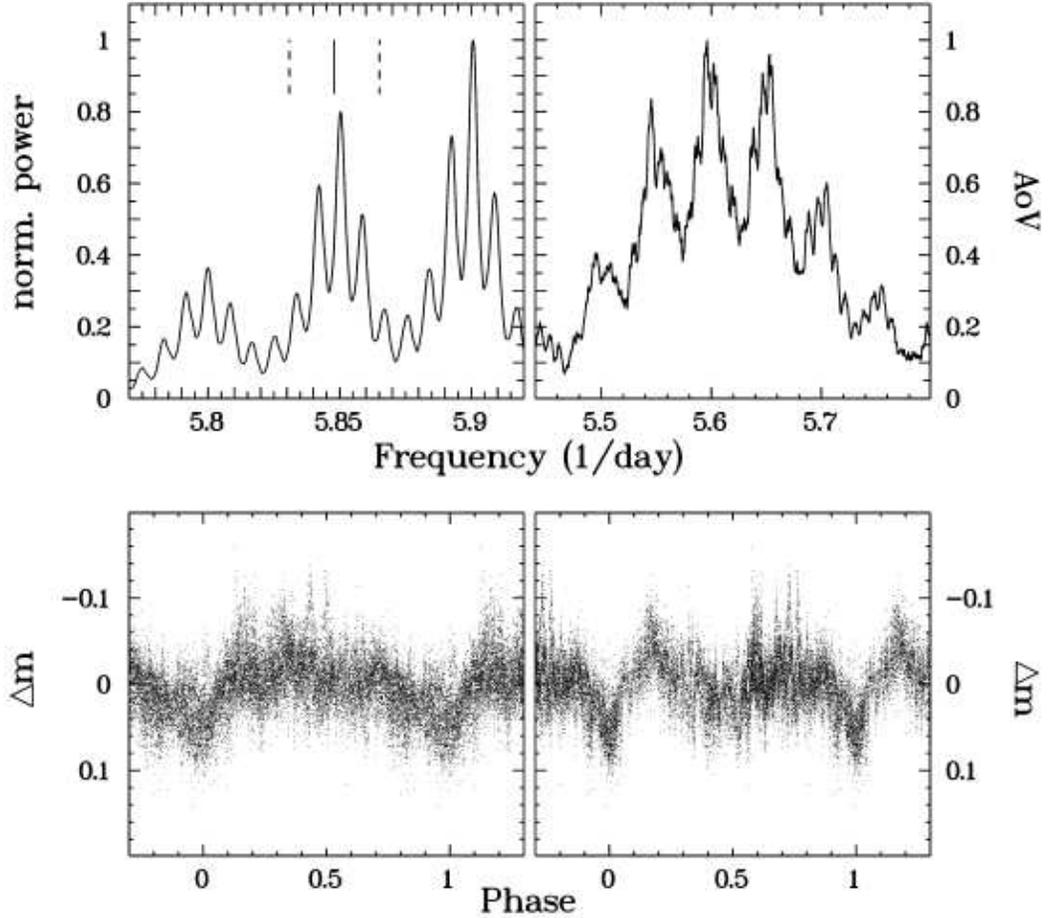}}
      \caption[]{{\em Upper left frame:} AoV
                 periodogram of the combined light curves of MU~Cen,
                 restricted to a narrow frequency
                 range close to 2 times the orbital frequency. The vertical
                 lines mark the frequency and the error limits corresponding 
                 to half the spectroscopic period. {\em Upper right frame}:
                 AoV periodogram of the residuals between the original light
                 curve and a best sine fit with a period fixed to the inverse
                 of the frequency of the
                 orbital peak in the AoV periodogram of the upper left 
                 frame. {\em Lower left frame:} Residual light curve folded on
                 the period $P_2$ (corresponding to the highest peak in the AoV
                 periodogram of the upper right frame). {\em Lower right frame:}
                 Residual light curve folded on period $2 \times P_2$.} 
\label{mucen-power}
\end{figure}

In order to verify if the brightness modulations are indeed coherent over
several nights, I first calculated a power spectrum of the combined light
curves of the four contiguous observing nights in June, using the
Lomb-Scargle algorithm (Lomb 1976, Scargle 1982, 
Horne \& Baliunas 1986). It shows a strong peak at a frequency very 
close to $2/P_{\rm spec}$ together with a nice pattern of daily aliases, 
leaving no doubt about the periodicity of the variations. Next, the combined
light curves of all data was subjected to a periodogram analysis. This
time, the analysis-of-variance (AoV) algorithm (Schwarzenberg-Czerny, 1989)
was preferred which leads to a higher contrast in the periodogram than the
Lomb-Scargle power spectrum. An expanded section of the frequency range
around $2/P_{\rm spec}$ is shown in the upper left frame of 
Fig.~\ref{mucen-power}. It is most 
interesting that the peak corresponding to $2/P_{\rm spec}$ [indicated together 
with the error range quoted by Friend et al.\ (1990) by the vertical 
tick marks] is only the second highest peak in the periodogram. Does this mean 
that the spectroscopic period is slightly in error? After all, it was determined
from observations spanning a short time base of only $\sim$52 hours or
6.4 cycles. As it will be shown below, this is not the case.

First, however, a sine curve with the period fixed to the period corresponding
to the peak close to $2/P_{\rm spec}$ in the AoV periodogram was fit to the
light curve. A second AoV periodogram was then calculated from the residuals 
between the original data and the best fit sine curve. In the upper right
frame of Fig.~\ref{mucen-power} a restricted frequency interval of this 
periodogram is shown. It contains several alias peaks. The highest of them 
corresponds to a period $P_2 = 0.178692\, {\rm days}$ which is slighly longer 
than $0.5 \times P_{\rm spec}$.
The minimum of this modulation can be expressed by the ephemeris:
\begin{displaymath}
{\rm BJD_{min}} = 2457182.495\, (18) + 0.178692\, (26) \times E
\end{displaymath}
where $E$ is the cycle number.
In view of the difficulty to define sound statistical error limits for periods
based on power spectra (or AoV periodograms) with complicated window functions
and in the presence of flickering 
(Schwarzenberg-Czerny 1991) only a conservative order 
of magnitude error (in brackets, expressed in units of the last significant 
digits of the period) is given, such that it would lead to an easily
recognizable phase shift of 0.1 over the total time base of the observations.
Similarly, the error of the epoch is simply assumed to be equal to 10\% of 
the period.

The folded light curve is shown in the lower left frame of 
Fig.~\ref{mucen-power}. The minimum has been chosen as phase 0. 
Flickering makes the light curve noisy, but there is no doubt about the
presence of consistent variations. The wave form is not sinusoidal, but 
rather exhibits a broad and structured asymmetrical maximum and a narrower 
minimum.

\begin{figure}
   \parbox[]{0.1cm}{\epsfxsize=14cm\epsfbox{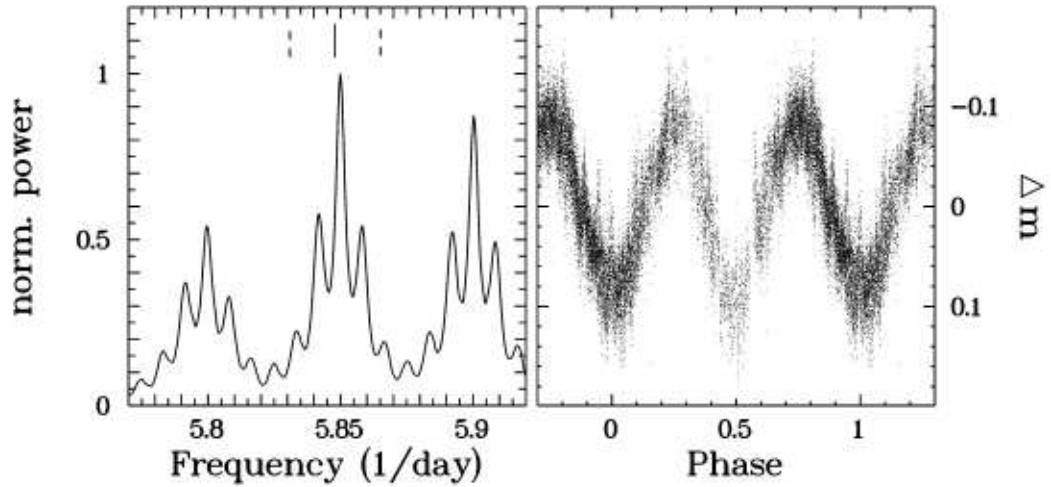}}
      \caption[]{{\em Left frame:} AoV periodogram of the combined 2015 light
                 curve of MU~Cen after subtraction of the modulations with
                 period $P_2$ (see text). The vertical
                 lines mark the frequency and the error limits corresponding 
                 to half the spectroscopic period. {\em Right frame:} Light
                 curve after subtraction of the modulations with period
                 $P_2$, folded on the refined orbital period, i.e., 
                 twice the period corresponding to the highest peak of the
                 AoV periodogram in the left frame.} 
\label{mucen-orbit}
\end{figure}

In order to remove
the strong scatter which is due to flickering, the curve folded on $P_2$ was 
binned into intervals of 0.01 in phase and subsequently smoothed by a cubic 
spline fit. This
modulation was then subtracted from the combined 2015 light curve which 
finally was again subjected to a period analysis. The corresponding
AoV periodogram is shown in the left frame of Fig.~\ref{mucen-orbit}. 
As in Fig.~\ref{mucen-power} twice the frequency of the spectroscopic 
variations and their error 
limits are shown as vertical tick marks. Now the highest peak is very close 
to $0.5 \times P_{\rm spec}$.

The much longer time base of the photometric observations, as compared to the
radial velocity curve of Friend et al.\ (1990), permits a refinement 
of the orbital ephemeris for MU~Cen. Interpreting the variations as being due 
to ellipsoidal variations of the secondary, the minima in the light curve should
coincide with the upper and lower conjunction between the cool star and the
white dwarf. The right frame of Fig.~\ref{mucen-orbit} shows the light curves 
folded on the refined period 
$P_{\rm orb} = 0.341883\, {\rm days}\, (8\hoch{h} 12\hochpunkt{m}3)$ 
deduced from the AoV periodogram. Light curve minima occur at:
\begin{displaymath}
{\rm BJD_{min}} = 2457182.459 \, (34) + 0.341883 \, (97) \times E
\end{displaymath}
Order of magnitude errors were determined in the same way as for $P_2$.

Although the minima have quite similar depth and thus do not permit an 
immediate decision whether these ephemeris refer to the upper or lower 
conjunction of the components of MU~Cen it will be shown in 
Sect.~\ref{Model fits} that the primary star is in front 
at phase 0.

The original light curve was investigated for further significant periodic 
modulations. However, apart from $P_{\rm orb}$ and $P_2$, none were found. 

\subsection{Aperiodic variations: Flickering}
\label{Aperiodic variations: Flickering}

It is well known that flickering is a hallmark of cataclysmic variables. It is
therefore not surprising that MU~Cen also exhibits this phenomenon (see 
Figs.~\ref{mucen-lc-1} and \ref{mucen-lc-2}), albeit with a relatively
low amplitude when compared to many other dwarf novae in quiescence.
An estimate of its amplitude can be derived from the difference
between the folded orbital light curve (right frame of Fig.~\ref{mucen-orbit})
and its smoothed version (Fig.~\ref{mucen-wdfit}) which will be used in
Sect.~\ref{Model fits} to adjust a model to the observed data.
A range of $\pm$3 of
the standard deviation of this difference may be regarded as the typical 
flickering amplitude: 
$A_{\rm f} = 0\hochpunkt{m}16$. This is significantly
lower than normally observed in quiescent dwarf novae. From the compilation
of {Beckemper (1995) an average amplitude of 
$A_{\rm f} = 0\hochpunkt{m}32 \pm 0\hochpunkt{m}17$ 
can be derived\footnote{Beckemper (1995)} used 
a somewhat different method to estimate the flickering amplitude than applied 
here. However, this does not invalidate the conclusion.}. 

\subsubsection{Power spectra}
\label{Power spectra}

In order to establish the quantitative properties of the flickering in MU~Cen,
I first investigate its frequency spectrum. As is well known, the power
spectrum of a light curve is a function of the frequency $\nu$ which is
proportional to the energy released by variations at that frequency. The
energy, in turn, is proportional to the square of the amplitude of the
underlying variation. Therefore, the root of the power spectrum can be
regarded as a measure of the amplitude of the variations 
in the light curve at frequency $\nu$ \cite{Bruch92}.

\begin{figure}
   \parbox[]{0.1cm}{\epsfxsize=14cm\epsfbox{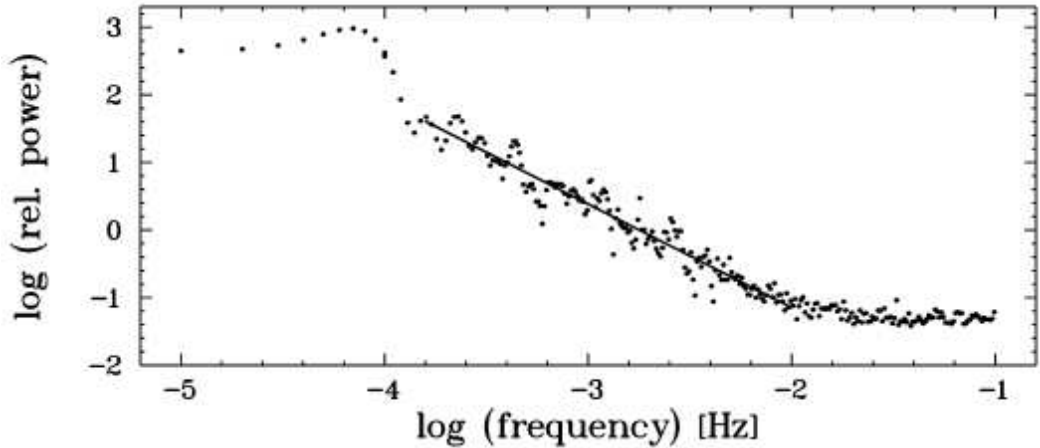}}
      \caption[]{Averaged power spectrum of the light curves of MU~Cen (dots).
                 The straight line is a least squares fit to the linear
                 part of the drop of power towards higher frequencies.}
\label{flickering-ps}
\end{figure}

On a double logarithmic scale the power spectra of all light curves of MU~Cen, 
calculated with the Lomb-Scargle algorithm, are remarkably similar. 
Therefore, their average, binned in phase at intervals of $\Delta \log(\nu)
= 0.01$ (frequency expressed in Hertz) was calculated. The result is shown
in Fig.~\ref{flickering-ps}. As is typically observed in CVs, the power
spectrum levels off at low frequencies. The hump close to $\log(\nu) = -4.1$
can be explained by the orbital variations discussed in 
Sect.~\ref{Periods in the light curve}. At higher frequencies the 
(logarithmic) power $P$ starts to drop linearly with an inclination $\alpha$, 
indicating that the flickering behaves like red noise ($P \sim \nu^\alpha$). 
At still higher frequencies ($\log(\nu) \ge -2$) the power spectrum again 
levels off, signaling the dominance of random measurement errors, manifest 
as white noise.

The slope of the linear drop determines the distribution of the relative
power of the flickering at different frequencies. It was measured by a least
squares fit of a straight line to the data in the interval 
$-3.83 \le \log(\nu) \le -2$ (solid line in Fig.~\ref{flickering-ps}). 
The resulting value 
$\alpha = -1.532 \pm 0.002$ 
is close to the upper end of the distribution of
corresponding values measured in other CV light curves by 
Bruch (1992)\footnote{Since Bruch (1992) measured $\alpha$ 
in the amplitude 
spectra, his values must be multiplied by two to be compared to the present 
data.} and Nolhen (1995) which span the range of 
$-2.64 \le \alpha \le -1.24$\footnote{disregarding the anomalous CV AE~Aqr.}.
The comparatively small value of $| \alpha |$ in MU~Cen indicates that 
the dominance of large over small amplitude variations in this system is 
less strong than in the majority of CVs. 

\subsubsection{Oscillations}
\label{Oscillations}

\begin{figure}
   \parbox[]{0.1cm}{\epsfxsize=14cm\epsfbox{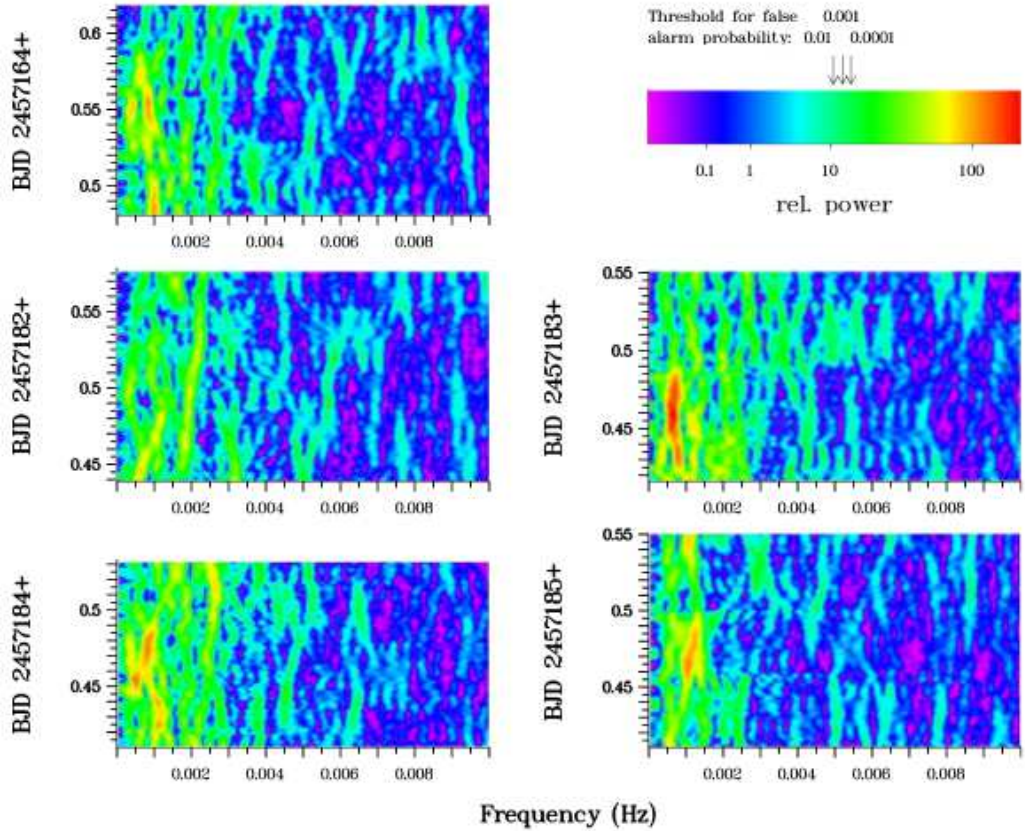}}
      \caption[]{Stacked power spectra of five light curves of MU~Cen.
                 Power is shown on a logarithmic scale as a function of
                 frequency and time. Note that structures in vertical
                 direction with an extend of less than $\sim$1 hour (about
                 one quarter of the vertical scale) are not independent. The
                 threshold values for false alarm probabilities 0.01, 0.001 
                 and 0.0001 are marked on the colour bar (top right).}
\label{stackedps}
\end{figure}

In order to verify if the individual events contributing to the flickering
are really independent, or if oscillations persistent in time are present,
stacked power spectra of the light curves of 2015, May and June\footnote{The
light curve of 2015, Feb.\ 11, is too short for this excercise.}
(discarding the data points before the gap in the observations on May 21;
Fig.~\ref{mucen-lc-2}) were calculated following Bruch (2014), and 
are displayed
in Fig.~\ref{stackedps}. Here, the power is shown on a logarithmic scale
as a function of frequency (horizontal axis) and time (vertical axis). 
Note that spectra within a time scale of 1 hour (that is approximately a
quarter of the total range in time in all frames) are not independent.
Therefore, only structures with a vertical extend of more than an hour
indicate persistent signals. 

The power levels corresponding to false alarm probabilities (i.e., the 
probability for a feature in the power spectra to be caused by noise) of
0.01, 0.001 and 0.0001 are indicated on the colour bar in the top right
panel of Fig.~\ref{stackedps}. They were calculated using eq.~18 of 
Scargle (1982) and depend strongly on the number $N_i$ of independent 
frequencies in the power spectrum. Horne \& Baliunas (1986) have
shown that $N_i$ is not only a function of the number of data points from
which the power spectrum is calculated, but also depends significantly on
the details of data sampling. Following Horne \& Baliunas (1986), $N_i$ was 
here estimated for each light curve
by generating 10\,000 data sets consisting of pure Gaussian noise, sampled
in exactly the same way as the real data. The distribution function of the
highest peaks in their power spectra\footnote{The power spectra were
oversampled in order to make sure that individual maxima are resolved and thus
an observed peak is in fact close to the maximum of the peak, and not just 
a point on the rising or falling branch.} was constructed and fit to the 
theoretical distribution curve (eq.~14 of Scargle, 1982) which 
contains $N_i$ as the only free parameter. Not
surprisingly, in view of the similarity of data sampling, $N_i$ is almost the 
same in the five nights considered here. Therefore, the threshold values 
for the false alarm probability are practically the same for all
power spectra show in Fig.~\ref{stackedps}.

Some intriguing features can be seen in the stacked power spectra. Maxima
which slowly evolve in strength and frequency can extend over the entire
duration of a light curve, suggesting the presence of persistent, albeit not
coherent oscillations. A particular clear-cut example is present on
2015, Jun 8 (second left hand frame in Fig.~\ref{stackedps}). During the
entire length of the light curve, the power spectra exhibit a maximum
which migrates slowly from 
$\sim$0.0016~Hz to $\sim$0.0023~Hz 
(or $\sim$10\hoch{m} to $\sim$7\hoch{m}). Similar features appear to be 
present also during other nights. Other strong signals have a temporal
extend of not significantly more than an hour and may therefore be due to
individual flickering flares.

The persistent power spectra features seen in some light curves of MU~Cen
share the characteristics attributed to quasi-periodic oscillations (QPOs)
sometimes seen in CVs. They occur on time scales of a couple of minutes, 
are not stable in frequency and are transient. Warner (2004) gives 
an overview of the rich phenomenology of such oscillations. The relationship 
between QPOs and flickering is not clear. Bruch (2014) pointed out 
that an accidental
superposition of random flickering flares, or a physical connection
between individual flares may well lead to their interpretation as QPOs.
This made him raise the question whether there is a conceptual difference
between the two phenomena at all.

\subsubsection{Wavelet analysis}
\label{Wavelet analysis}

The wavelet transform (see Jawerth \& Sweldens, 1994, Chui, 
1992, and Scargle et al., 1993, for general introductions) 
is an alternative to the Fourier transform (and thus power spectra) 
when analyzing
stochastic data such as flickering. It permits the decomposition of a signal 
according to a localized function, the (finite) carrier of which is tied to 
the investigated scale. The base functions -- the wavelets -- are scaled 
versions of a fundamental function, the mother wavelet. The wavelet transform
of a time-dependent signal is then a representation of the signal in time and 
frequency.

The application of wavelet techniques to flickering light curves was
pioneered by Fritz \& Bruch (1998). They showed that the scalegram
\cite{Scargle93}, which basically measures the variance of the
wavelet coefficients and thus of the modulated part of the signal (e.g.,
a lightcurve) as a function of the time scale, permits, when suitable 
normalized, a direct comparison between light curves
observed under different conditions. They also found that the scalegrams
of flickering light curves on a double logarithmic scale always exhibit
a linear rise from small to long time scales, when disregarding very
small (data noise) or very long (variations not representing flickering)
time scales. This enables to condense the essence of the scalegram
into two parameters, the slope $\overline{\alpha}$\footnote{Here, the symbol 
$\overline{\alpha}$ is used in order to avoid confusion with $\alpha$ 
introduced earlier as the inclination of the linear part of the flickering
power spectrum on a double logarithmic scale.} of a straight line fitted to 
the scalegram points, and $\Sigma = \log S(t_{\rm ref})$, where S is the 
scalegram value and $t_{\rm ref}$ is a reference timescale 
(for details, see Fritz \& Bruch, 1998). $\overline{\alpha}$ measures 
the distribution of the strength of the flickering among different
timescales, while $\Sigma$ measures the overall strength of the flickering 
with respect to the unmodulated background light.

   \begin{figure}
   \parbox[]{0.1cm}{\epsfxsize=14cm\epsfbox{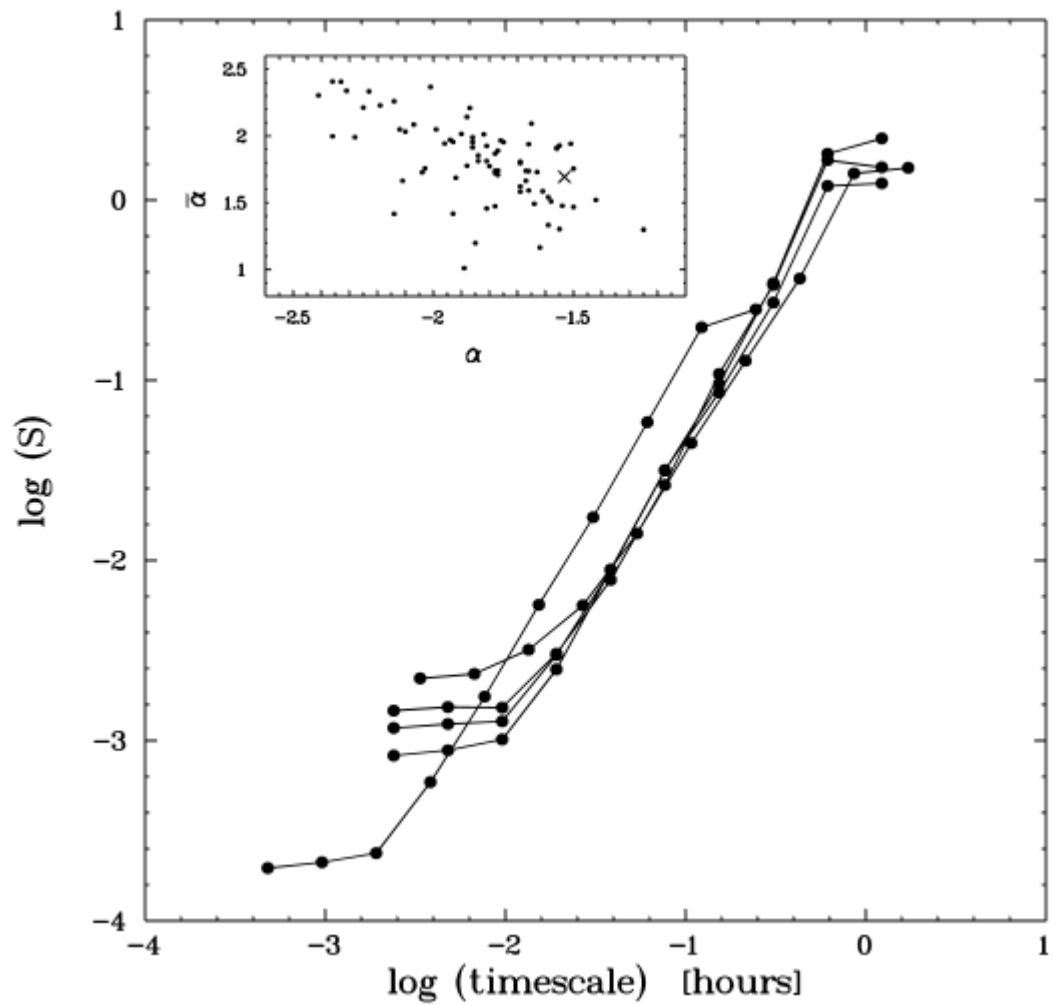}}
      \caption[]{Energy normalized scalegrams of the 1995 May and June
                 light curves of MU~Cen. The insert shows the relation
                 between the inclination $\alpha$ of the linear part of
                 the double logarithmic power spectra of flickering light
                 curves (from Nohlen, 1995) and the scalegram parameter
                 $\overline{\alpha}$ (from Fritz \& Bruch, 1998). The
                 location of MU~Cen is indicated by a cross.}
\label{scalegram}
\end{figure}

The normalized scalegrams of the 2015 May and June light curves of MU~Cen 
are shown in Fig.~\ref{scalegram}. They are remarkably similar to each
other. Even the one deviating graph, corresponding to the light curve of
June 9, differs only slightly from the others when compared to the much
larger scatter found in different light curves of the same object by
Fritz \& Bruch (1998). Consequently, the parameters $\overline{\alpha}$
and $\Sigma$ [choosing 3 minutes as reference time scale as has been done by
Fritz \& Bruch (1998)] are exceptionally well defined: 
$\overline{\alpha} = 1.70 \pm 0.10$ and 
$\Sigma = -1.76 \pm 0.24$. 
This confines MU~Cen to
a small range of the $\overline{\alpha} - \Sigma$ diagram at the lower
edge of the region occupied by quiescent dwarf novae
(see Figs.~12 -- 14 of Fritz \& Bruch, 1998). 

The small (for a quiescent dwarf nova) value of $\Sigma$ reflects the
comparatively low amplitude scale of the flickering in MU~Cen. 
$\overline{\alpha}$, measuring the distribution of the strength of the
flickering on different time scales, should not be independent of the
inclination $\alpha$ of the linear part of the double logarithmic power
spectrum. This is confirmed by the inset in Fig.~\ref{scalegram}, where
$\overline{\alpha}$ is plotted as a function of $\alpha$ for all white 
light and $B$-band light curves studied by both, Fritz \& Bruch (1998)
and Nohlen (1995). In spite of considerable scatter both entities 
are clearly correlated. The location of MU~Cen is indicated by a cross.

\section{Model fits}
\label{Model fits}

In order to delimit some system parameters of MU~Cen, I will compare the 
observed 
light curve with results of some simplified model calculations. To this end,
I employ the Wilson-Devinney code (Wilson \& Devinney, 1971, Wilson, 1979), as
implemented in MIRA, in
the mode appropriate for semi-detached binaries. It is not the aim of this
exercise to model the MU~Cen system in detail, but rather to see if the
ellipsoidal variations can be reproduced with reasonable system parameters. 
Therefore, the inadequacy of the Wilson-Devinney code to describe the primary
component of a CV (white dwarf, accretion disk, hot spot) does not invalidate
the results. As will be seen, a good fit to the data can be achieved with 
the primary modeled as a moderately hot constant star, influencing the
shape of the observed variations only through reflection off the secondary
star. Of course, this is not a physical description of the primary, but just
a parameterization.

The Wilson-Devinney model involves many parameters. Of these, the following
were adjusted to the data: the mass ratio $q = M_2/M_1$ of the components, the
orbital inclination $i$, and, for the secondary star, the temperature $T_2$, 
the limb darkening coefficient $u_2$, the gravity darkening coefficient $g_2$ 
and the albedo $\cal{A}$$_2$.
The mass ratio also determines the potential at the Roche surface and thus
the radius of the secondary star in units of the component separation.
The primary component was parameterized by its temperature and its surface
potential (determining its size in the Roche geometry). Moreover, a phase 
shift was taken into account in order to make up for a possible error in the 
ephemeris. Finally, the normalization constant was also adjusted.

The isophotal wavelength $\lambda_{\rm iso}$ of the observed bandpass 
has only a small influence on the 
results and was therefore fixed. In order to estimate $\lambda_{\rm iso}$, 
I assume the telescope and atmospheric transmission functions to be constant 
over the range of sensitivity of the detector. A rough stellar energy 
distribution function can be obtained assuming the ratio of the radiated energy 
of the secondary star and the primary component (only considering the 
contribution of the accretion disk) to be equal to 2.75 in the observed 
bandpass (see below). The cool star is assumed to radiate like a black body 
with a temperature of 5\,000~K (see below). For the 
primary a standard steady state accretion disk spectrum is 
assumed\footnote{This is not realistic because a dwarf nova in quiescence
is not expected to have a steady state accretion disk. However, in view of
all the other approximations made here this assumption should be sufficient
for a rough estimate of the isophotal wavelength of the light curve data; 
the more so considering the small dependence of the model fits on 
$\lambda_{\rm iso}$.},
calculated for a mass transfer rate of $10^{-9}\, M_\odot/{\rm year}$, a primary
star mass of $M_1 = 0.8\,M_\odot$, an inner disk radius corresponding to the 
radius of a white dwarf with that mass, and an outer disk radius of $0.4 \, A$, 
where $A$
is the component separation calculated from $P_{\rm orb}$, $M_1$ and $q=0.83$ 
(Friend et al., 1990), using Kepler's law. This 
approximation to the spectral energy distribution
of MU~Cen, together with the sensitivity function of the CCD detector used
for the observations, then yields an isophotal wavelength of $\sim$5960~{\AA}.

\begin{figure}
   \parbox[]{0.1cm}{\epsfxsize=14cm\epsfbox{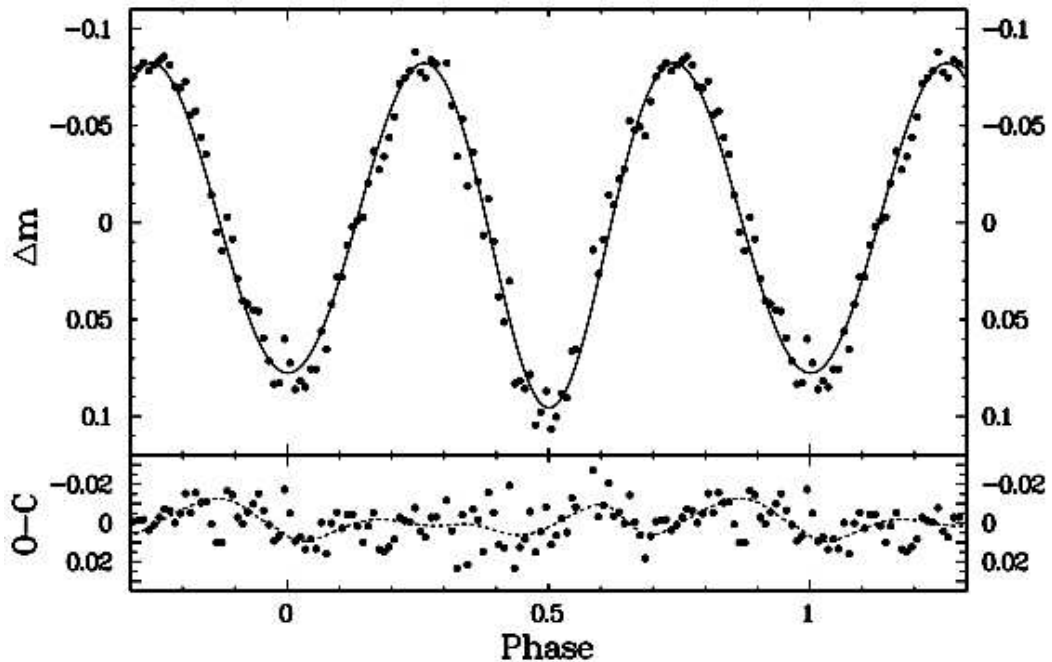}}
      \caption[]{{\em Upper frame:} Phase folded and binned light curve of
                 MU~Cen (dots) and best fit model light curve (solid line).
                 {\em Lower frame:} Difference between observed and model
                 light curve.}
\label{mucen-wdfit}
\end{figure}

In order to fit the model to the data, the phase folded light curve of MU~Cen
was binned into 100 phase bins. Three of them were rejected because they were
strongly contaminated by flickering flares which were not sufficiently averaged
out after phase folding. The resulting average light curve is shown as dots in 
the upper frame of Fig.~\ref{mucen-wdfit}. 

In contrast to Fig.~\ref{mucen-orbit}, it now is obvious that the minimum at
phase 0.5 is slightly deeper than that at phase 0. This is expected if phase
0 correponds to the upper conjunction of the secondary star because then the
observer sees more of the side illuminated by the primary than during lower
conjunction. 

The procedure to find the best fit of the model to the data consists in 
minimizing the $\chi^2$ of the residuals between model and data. This is not
an easy task in view of the numerous adjustable parameters. Applying standard
methods such as the SIMPLEX algorithm (Caceci \& Cacheris, 1984) it is
quite likely to get stuck in a local minimum of the multidimensional 
$\chi^2$ surface instead of finding the global minimum. Therefore, a
statistical approach is preferred here.

To this end, 1225 trial calculations were performed. In each case initial 
values for the variable parameters were selected at random within an ample
range of plausible choices. These were used to derive a start value for 
$\chi^2$. Then each parameter was changed slightly, keeping all others 
constant, and $\chi^2$ was calculated again. If this yielded a $\chi^2$ 
smaller than the start value, the parameter set was adjusted to that which
resulted in the smallest $\chi^2$ and the entire process was repeated until no
further improvement was possible. In this way a parameter set was found which
corresponds to the local minimum of the $\chi^2$-surface nearest to the 
$\chi^2$ derived using the initial parameters.  

\begin{figure}
   \parbox[]{0.1cm}{\epsfxsize=14cm\epsfbox{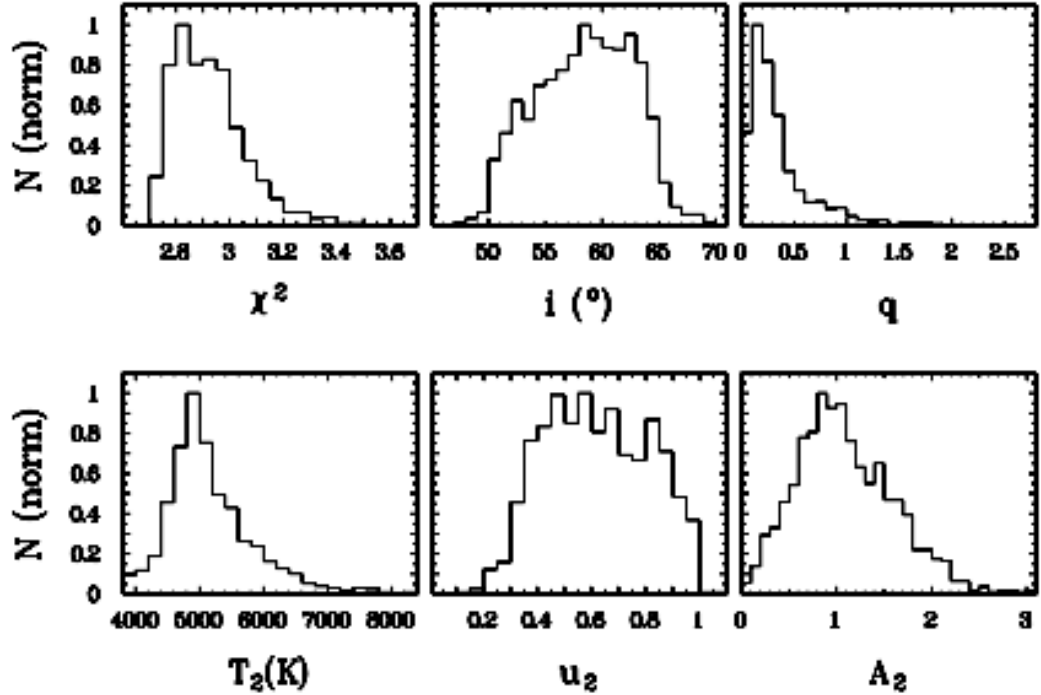}}
      \caption[]{Normalized distribution of $\chi^2$ and of the best fit 
                 values for the
                 most relevant system parameters (orbital inclination $i$, 
                 mass ratio $q=M_2/M_1$, secondary star temperature $T_2$,
                 limb darkening coefficient of secondary $u_2$, and secondary
                 star albedo $\cal{A}$$_2$), constructed from the results of
                 numerous model fits. See text for details.}
\label{mucen-pardist}
\end{figure}

The distribution of the minimal $\chi^2$ and of the most relevant system
parameters derived from the 1225 trials is shown in Fig.~\ref{mucen-pardist}.
The parameter set corresponding the smallest $\chi^2$ encountered in this
exercise was used to calculated the model light curve shown as a solid line
superposed upon the observational data in the upper frame or 
Fig.~\ref{mucen-wdfit}. The difference between data and model results, 
i.e., the $O-C$ curve, is
plotted in the lower frame of the figure. While the overall model fit to the
data appears satisfactory, the residuals are not completely random, indicating
that the model does not describe the observations perfectly as may be
expected in view of the simple parameterization of the primary component which
cannot reflect its true structure. Indeed, $R$ statistics 
(Bruch, 1999) reveal a probability of $98\%$ that the $O-C$  
curve contains correlated variations.

In order to obtain a feeling for the goodness of the fit
a bootstrap analysis was performed. To this end, 
a smoothed version of the $O-C$ curve was first generated (broken line in
the lower frame of Fig.~\ref{mucen-wdfit}). The probability of correlated
variations in the difference between the smoothed and the original $O-C$
curve (which may be termed ``second order $O-C$ curve'') is significantly
reduced to 57\%. The smoothed $O-C$ curve is added to the best fit model
light curve which is thus corrected for systematic errors of the model.
This light curve is then subjected to random noise with a Gaussian 
distribution and a variance equal to the variance of the second order 
$O-C$ curve, generating thus a simulated observational light curve. Finally,
the $\chi^2$ between this curve and the best fit model light curve is 
calculated. This procedure was repeated 1\,000 times. The distribution of 
the individual $\chi^2$ values then yields a statistical error 
$\sigma_{\chi^2} = 0.40$.

The total range of the $\chi^2$ distribution shown in the upper left frame of
Fig.~\ref{mucen-pardist} is 
0.82. This is only $2.05 \sigma_{\chi^2}$. 
The global minimum of the $\chi^2$ hypersurface is therefore not expected to 
be much deeper than any of the local minima found in the trial calculations
because otherwise a larger dispersion of the individual $\chi^2$ values should 
be observed. Thus, all of the
trial fits which led to this distribution are statistically acceptable.
Consequently, the same is true for the model parameters shown in the figure,
although in this case physical considerations may constrain some of them.

One of these cases is the albedo $\cal{A}$$_2$ of the secondary star. The
distribution peaks close to $\cal{A}$$_2 = 1$ (lower right frame of
Fig.~\ref{mucen-pardist}; the median values is 
$\cal{A}$$_2 = 1.0$), 
but has a long tail to values $>$1. Since 
re-distribution of radiation incident upon the secondary star from outside 
the rather broad observed wavelength range at a scale leading to an apparent 
albedo of up to 3 is quite unlikely, values of $\cal{A}$$_2$ significantly
larger than 1 may be 
considered unphysical. Nevertheless, the model results indicate that the
albedo of the MU~Cen secondary is high. Rafert \& Twigg (1980) have shown that 
on average ${\cal A} \approx 1$ and 
${\cal A} \approx 0.5$ for stars with radiative and convective envelopes, 
respectively. The assumed temperature of the MU~Cen secondary should place 
it into the latter category, contrary to what is found here. 
However, the albedo of individual stars can deviate strongly from the 
average. At least in one case (RX~Ari; a detached binary star) 
Rafert \& Twigg (1980) find ${\cal A} \approx 1$ for a star with a 
considerably lower temperature than the MU~Cen 
secondary\footnote{Rafert \& Twigg (1980) speculate that this result may be 
artificial due to 
the possible presence of a hot spot caused by the impact of mass transferred 
from the second star in the system.}. 

For the limb darkening of the secondary star anything more sophisticated 
than a simple linear law of the
kind $I(\mu)/I(1) = 1 - u (1-\mu)$ is not justified in 
view of the uncertainties inherent in the present approach. Here, $I(1)$ is the
specific intensity at the centre of the stellar disk, and $\mu = \cos \gamma$,
where $\gamma$ is the angle between the line of sight and the emergent
radiation. The distribution of $u_2$ (Fig.~\ref{mucen-pardist})
is rather broad, with a median value of 
$u_2 = 0.61$. 
This is consistent
with limb darkening coefficients interpolated in the tables of
Claret (2004) at the appropriate values for the effective temperature and
wavelength. Likewise, a distribution of the gravity darkening coefficient
of the MU~Cen (not shown) has a median value of 
0.57, 
which is consistent
with the values found in the tables of Claret \& Bloemen (2011).

The distribution of the best fit values of the temperature $T_2$ of the
secondary star has a strong peak just below 5000~K. The median is 5017~K.
This is consistent with the temperatures of the secondaries of other CVs 
with similar orbital period. 
Table~\ref{Parameters of long period CVs} contains all CVs identified
in the most recent on-line version of the Ritter \& Kolb catalogue
(Ritter \& Kolb, 2003) with a period within 0.04 days of that of MU~Cen 
for which relevant data could be found in the literature. Only for two of
them direct quotes of the secondary temperature exist (shown in italics
in the forth column of the table) but for many more the spectral type has
been determined. In order to transform spectral types into temperatures, the 
average temperature of luminosity class V stars from Table~A1 of 
Cenarro et al.\ (2007) was adopted, and the results are given in 
Table~\ref{Parameters of long period CVs}. When only a range of
types for the CV secondary is quoted in the literature, the temperature
corresponding to the centre of this range is chosen. From the scatter of the 
individual temperatures values at a given spectral type an error of 
$\sim$150~K is estimated. The secondary stars in long period CVs are expected 
to be slightly evolved. Therefore, the assumption of the temperature of a 
main sequence star of the same
spectral type introduces a systematic error. In order to estimate its order
of magnitude, the average temperature difference between luminosity class V 
and IV stars in the relevant range of spectral types (G7 - M0) was determined
to be $\sim$340~K from the list of Cenarro et al.\ (2007). Thus, the 
temperatures quoted in Table~\ref{Parameters of long period CVs} may be 
slightly on the high side. Table~\ref{Parameters of long period CVs} does not 
contain any indications of a systematic variation of the secondary temperature 
within the small range of orbital periods sampled. This is not surprising. 
Using a smaller data sample, Beuermann et al.\ (1998) have 
already shown the significant scatter
of secondary spectral types at a given period for periods above 
$\sim$5 hours. The average of the temperature values quoted in 
Table~\ref{Parameters of long period CVs} is $4670 \pm 400 \, {\rm K}$. Within
the error limits this is equal to the median of the temperature distribution 
found for MU~Cen.

\begin{table*}
\caption{Parameters of long period CVs}
\label{Parameters of long period CVs}

\hspace{1ex}

\begin{tabular}{lccccl}
\hline
Name      & Period      & Spectral    & Temp.  & Mass     & Ref. \\
          & (days)      & tpye        & (sec.) & ratio    &            \\
          &             & (sec.) & (K)          & $M_2/M_1$ &           \\
\hline
SY Cna    & 0.38238     & G8          & 5280         & 1.18     & 3          \\
IU Leo    & 0.37631     & K4.5        & 4490         & 0.85     & 13         \\
AT Ara    & 0.37551     & K2          & 4840         & 0.79     & 1          \\
RU Peg    & 0.3746\phantom{0}& K5V    & {\it 4600}   & 0.88     & 4,9        \\
CSS0467+49& 0.37155     & K4.5        & 4490         & 0.73     & 14         \\
QZ Aur    & 0.35750     &             & {\it 5200}   &          & 2          \\
GY Hya    & 0.34724     & K4-K5       & 4490         &          & 6          \\
CH UMa    & 0.34318     & K5.5        & 4400         & 0.43     & 12         \\
BT Mon    & 0.33381     & G8V         & 5280         & 0.84     & 7          \\
V1309 Ori & 0.33261     & K6-K8       & 4130         & 0.55     & 5,8        \\
V392 Hya  & 0.32495     & K5-K6       & 4400         & 0.55     & 6          \\
AF Cam    & 0.32408     & K4-M0       & 4140         & 0.60     & 11,13      \\
V363 Aur  & 0.32124     & G7          & 5280         & 1.17     & 10         \\
IPHAS J034511.59+533514.5 & 0.31390    & K5.5        & 4400         & 0.83     
          & 13         \\
RY Ser    & 0.3009\phantom{0} & K5    & 4540         & 0.84     & 12         \\
AC Cnc    & 0.30048     & K2V         & 4840         & 1.02     & 10         \\
\hline
\end{tabular}

{\bf References:} 
(1) Bruch (2003); 
(2) Campbell \& Shafter (1995); 
(3) Casares et al.\ (2009); 
(4) Dunford et al.\ (2012); 
(5) Howell et al.\ (2010);
(6) Peters \& Thorstensen (2005);
(7) Smith et al.\ (1998);
(8) Staude et al.\ (2001);
(9) Stover (1981); 
(10) Thoroughgood et al.\ (2004);
(11) Thorstensen \& Taylor (2001); 
(12) Thorstensen et al.\ (2004); 
(13) Thorstensen et al.\ (2010);
(14) Thorstensen \& Skinner (2012)

\end{table*}

Disregarding the faint outer wings of the distribution, the best fit values for
the orbital inclination $i$ range between 50\hoch{o} and 65\hoch{o} and are
somewhat skewed to the upper half of this range. The median value is 
58\hochpunkt{o}6. 
This is compatible with the absence of eclipses in MU~Cen.

Most surprising is the distribution of the best fit values of the mass ratio
$q = M_2/M_1$ (upper right frame of Fig.~\ref{mucen-pardist}). From the radial
velocity amplitude in the infrared region, combined with measurements of the
rotational line broadening, Friend et al.\ (1990) determine 
$q = 0.83 \pm 0.16$
for MU~Cen. This is quite close to the average ($0.80 \pm 0.22$) of the
literature values of $q$ for CVs with an orbital period close to that of 
MU~Cen as quoted in the fifth column of 
Table~\ref{Parameters of long period CVs}. In contrast, the distribution
shown in the figure strongly peaks at quite small values of $q$ and has a
long tail to large values, even reaching values much above $q=1$. 
The latter may be considered unphysical, 
meaning that the secondary is much more massive that the primary which
would result in unstable mass transfer. But may $q$ really be
much smaller than measured by Friend et al.\ (1990) and expected 
for CVs with similar orbital periods?  

\begin{figure}
   \parbox[]{0.1cm}{\epsfxsize=14cm\epsfbox{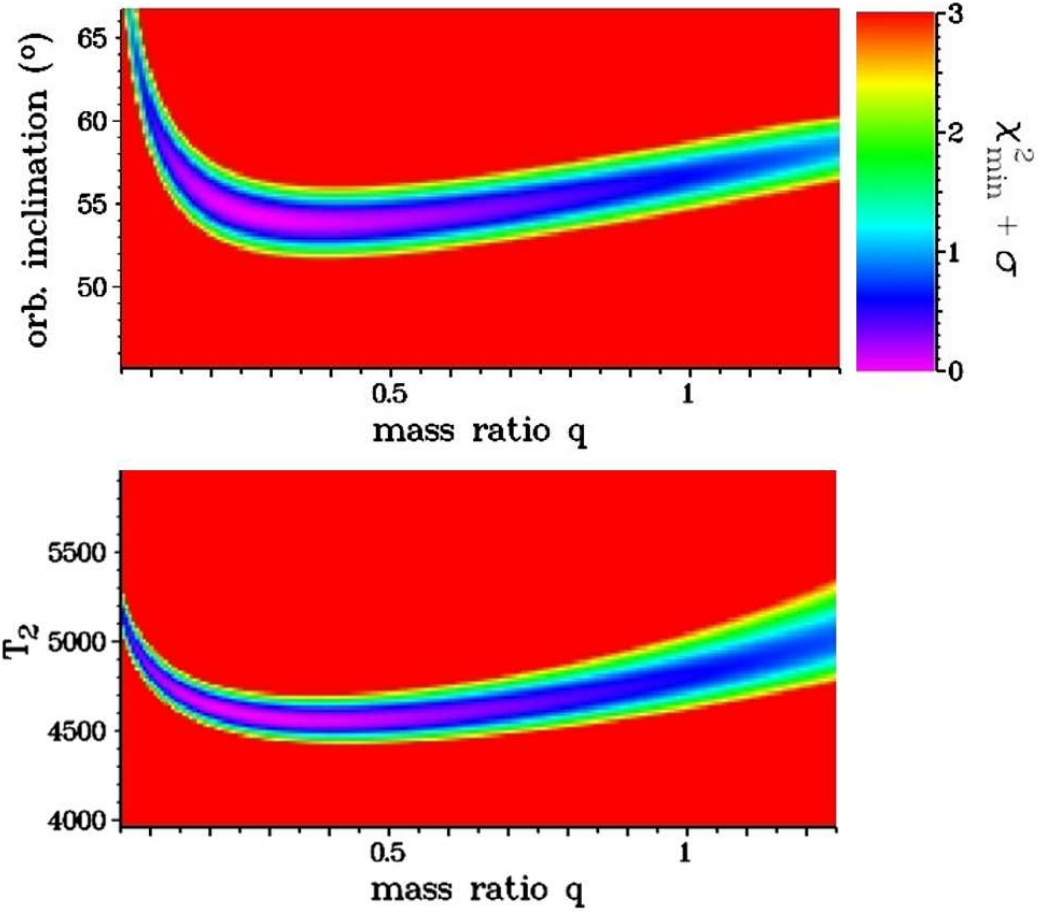}}
      \caption[]{Projection of the $\chi^2$ hyperspace on the $q - i$ and 
                 the $q - T_2$ plane, respectively, with all the other 
                 variable parameters fixed to the best fit values. The 
                 colour code is such that purple refers to $\chi^2_{\rm min}$, 
                 the smallest value encountered in the plane, and red to 
                 $\chi^2 \ge \chi^2_{\rm min} + 3 \sigma_{\chi^2}$. }
\label{mucen-parcorr}
\end{figure}

This is quite unlikely. Fig.~\ref{mucen-parcorr} shows the projection of the
$\chi^2$ hyperspace on the $q - i$ and the $q - T_2$ plane,
respectively, with all the other variable parameters fixed to the best fit
values. The colour code is such that purple refers to $\chi^2_{\rm min}$, 
the smallest value encountered in the plane, and red to 
$\chi^2 \ge \chi^2_{\rm min} + 3 \sigma_{\chi^2}$. It is obvious that for a wide
range of mass ratios model solutions can be found which lead to $\chi^2$ 
within one standard deviation of the best fit value. 
These results are not restricted to the projections shown in the figure, but
repeat themselves in other projections involving $q$. This means
that the model fits are unable to reliably restrict the mass ratio. 

Finally, the model calculations permit to quantify the relative contribution
of the secondary star to the total light of MU~Cen.
The best fit model solution implies that on the average over the orbital
period the secondary star is 
2.75 
times brighter than the primary in the observed wavelength range. 

\section{Discussion}
\label{MU Cen Discussion}

\subsection{The period $P_2$}
\label{The period P2}

The light curve of MU~Cen was found to contain consistent modulations on two
different periods. Apart from the dominating orbital period $P_{\rm orb}$
which is due to ellipsoidal variations of the secondary star, variability on a 
second period $P_2$, slightly longer than $0.5 \times P_{\rm orb}$ was
detected. There is no obvious simple 
relation between $P_2$ and $P_{\rm orb}$.  The total amplitude (discounting
the scatter due to flickering) of the $P_2$ modulation is approximately 
0\hochpunkt{m}07, while
the total orbital amplitude of $\approx 0\hochpunkt{m}20$ is almost three
times as high. The nature of the $P_2$ variations is not immediately obvious. 

One may speculate
that MU~Cen is an intermediate polar. In that case the modulation may be
due to the variable aspect of a magnetically confined accretion region
on the surface of a white dwarf rotating with $P_2$. 
However, there are no further convincing indications for an intermediate polar
nature of MU~Cen. One of the telltale signatures of such stars are high 
excitation emission lines in their spectra. In this respect the literature 
contains ambiguous information: while Warner (1976) mentions the 
``possible'' presence of He~II in the quiescent spectrum, it is definitely 
absent in the spectrum shown by Zwitter \& Munari (1996), also 
taken during quiescence. The ratio between orbital and spin period 
($P_{\rm orb}/P_{\rm rot} = 1.91$) would be rather unusual -- even if not 
unprecedented -- for intermediate polars. Mukai's 
list\footnote{http://asd.gsfc.nasa.gov/Koji.Mukai/iphome/iphome.hmtl}
contains only four confirmed systems with $P_{\rm orb}/P_{\rm rot} < 2.5$.  
Moreover, intermediate polars are generally seen in x-rays. But the VizieR 
data base\footnote{http://vizier.u-strasb.fr/index.gml} does not contain 
an x-ray 
source which might be associated with MU~Cen. Thus, the evidence for MU~Cen 
to be an intermediate polar is so meager that this hypothesis can 
probably be discarded. 

Alternatively, $P_2$ may be interpreted as the first overtone of a signal
with twice that period. The light curve, folded on 
$2 \times P_2 = 0.3574$~days is shown in the lower right frame of
Fig.~\ref{mucen-power}. Not surprisingly, just as in the light curve
folded on $P_2$, a consistent modulation is apparent; in this case a double
humped structure, embraced by a narrow minimum which almost looks like a 
shallow eclipse. The excess of $2 \times P_2$ over $P_{\rm orb}$ is
$\epsilon = \left( 2\times P_2 - P_{\rm orb} \right) / P_{\rm orb} 
= 0.045$ which
is reminiscent of the period excess typically observed for superhumps.
Such features were originally thought to be restricted to superoutbursts
of SU~UMa type dwarf novae. However, they are increasingly found also in
longer period CVs. The on-line version of the 
Ritter \& Kolb (2003) catalogue lists 14 systems 
with periods above the period gap exhibiting superhumps, although none of
them has a period as long as MU~Cen\footnote{Variations at 
$P=0\hochpunkt{d}494027$
observed in EY~Cyg ($P_{\rm orb}=0\hochpunkt{d}45932448$) by
Costero et al.\ (2004) are not confirmed as being due to superhumps.}.
Moreover, superhumps are not restricted to accretion disks 
in their bright state such as dwarf nova in outburst, novalike variables or
old novae (Osaki \& Kato, 2014). On the other hand,
the mass ratio of CVs on average increases with the orbital period. As a 
consequence of the growth of the superhump period excess with the mass ratio 
(see, e.g., Smith et al., 2007), $\epsilon$ is then also expected to 
be larger
for long period systems. The observed period excess of MU~Cen is much smaller
than anticipated if the variations were due to a superhump. Furthermore,
the shape of the variations (Fig.~\ref{mucen-power}) is quite different
from a normal superhump profile.  

Thus, I cannot offer a convincing explanation for the origin of the $P_2$ 
variations seen in MU~Cen and this issue remains open.

\subsection{Spectrum vs photometry: a contradicion?}
\label{Spectrum vs photometry: a contradicion?}

In order for the orbital light curve to be as strongly dominated by 
ellipsoidal variations as they appear to be one would expect the cool
star to leave a strong imprint on the spectrum. However, this seems to
be slightly at odds with the available observations. The only published
spectrum (Zwitter \& Munari, 1996) shows indeed the Na-D lines and 
possibly other atomic lines in absorption, but neither the line spectrum nor the
continuum in the red range exhibit as strong a contribution of a late
type star as would be expected in view of the photometric variations.

According to the long term light curves of MU~Cen provided by the AFOEV and
AAVSO on their respective web-sites, the spectrum of MU~Cen was taken at an
epoch well separated from the previous and subsequent outbursts. However, 
on the exact date the AAVSO light curve places MU~Cen at a visual magnitude
of 14.2. This is 0\hochpunkt{m}6 above the mean quiescent $V$ magnitude
(Bruch \& Engel, 1994) which leaves some margin to reconcile the 
spectroscopic findings with the present photometric results.

As mentioned in Sect.~\ref{Model fits} the secondary star was 
found to be 2.75 times brighter than the primary component. 
If the magnitude of MU~Cen at the epochs
of the present observations was equal to the mean quiescent magnitude of 
14\hochpunkt{m}78 (Bruch \& Engel, 1994), and if the elevated 
brightness of  14\hochpunkt{m}2 at the epoch of the spectroscopic observations 
of Zwitter \& Munari (1996) was due to a brighter primary component, 
a flux ratio of 
$F_{\rm sec}/F_{\rm prim} = 0.75$ 
is calculated for that 
epoch. This may explain why the signature of the secondary star is not as 
strong in the observed spectrum as might be expected considering its dominance 
in the light curve.

\subsection{Flickering behaviour}
\label{Flickering behaviour}

One of the initial purposes to observe MU~Cen was the verifications of the 
hitherto not confirmed presence of flickering and the study of its 
properties. As expected,
MU~Cen indeed does flicker, however on a comparatively low level
compared to most CVs investigated by Beckemper (1995).
This can be explained by the strong
contribution of the secondary star to the total light. If the secondary
star radiation were negligible, as it is in most of the stars of the sample
of Beckemper (1995), the flickering amplitude would raise to 
$0\hochpunkt{m}62$, 
even higher than the average value for quiescent dwarf novae and close to 
the amplitude observed in strongly flickering systems such as WW~Cet and 
WX~Hyi (Beckemper 1995).

The frequency behaviour of the flickering is not unusual when compared to
large samples of other CVs. The red noise characteristics, as parameterized 
by the slope $\alpha$ of the power spectrum on a double logarithmic scale,
is within the range observed otherwise, albeit close to its borders. 
Similarly, the scalegram parameters $\overline{\alpha}$ and $\Sigma$, derived 
from a wavelet analysis of the flickering, place MU~Cen within the range
occupied by other quiescent dwarf novae. The location close to the lower
edge of that range is explained by the dilution of the flickering light 
source by the strong contribution of the secondary star's light. 
Night-to-night variations of $\overline{\alpha}$ and $\Sigma$ are much
smaller than observed in most CVs (Fritz \& Bruch, 1998).

Interestingly, the stacked power spectra of MU~Cen contain indications that
flickering events do not occur quite at random but may be correlated,
leading to (incoherent) oscillations with slowly evolving amplitudes and
frequencies. This may blurr the borderline of flickering and what is termed
quasi-periodic oscillations in the literature. It may be worthwhile to 
submit a larger sample of CVs to a systematic study of stacked power spectra 
of their light curves.  

\subsection{System parameters}
\label{System parameters}

The attempt to pin down system parameters by modelling the light curve was
only met with limited success. The inadequacy of the Wilson-Devinney model
to describe the primary component of a cataclysmic variable is not considered
to be the decisive factor, as long as one refrains from drawing conclusions 
on the primary's detailed physical properties. I rather attribute the 
limitations met by the model calculations to the scatter in
the phase folded light curve, remaining even after binning as a reflection
of the influence of flickering which is not completely cancelled out in the
average curve. As a result, many combinations of model parameters lead
to calculated light curves which are statistically compatible with the
observed one. This inhibits their precise determination.

The model fits permitted to determine only two important system parameters
with reasonable certainty. The orbital inclination should lie in the range
$50^o \le i \le 65^o$, compatible with the absence of eclipses. A formal
Gauss fit to the distribution shown in the lower left frame of 
Fig.~\ref{mucen-pardist} yields $4950 \pm 380$~K for the temperature of the 
secondary star, slighly lower than the median value of the distribution
cited in Sect.~\ref{Model fits}. This is in line with secondary star
temperatures found in other CVs with similar orbital periods.
The less important parameters limb and gravity darkening coeffients are 
in line with expectations. In contrast, the albedo of the secondary star
appears to be higher than than expected for a star with a convective
envelope. 
Most unfortunate is the inability to derive an independent estimate for the
mass ratio $q$ of MU~Cen. Strong correlations with other parameters permit to 
find parameter combinations which lead to statistically acceptable model fits
for a very wide range of $q$. 

Friend et al.\ (1990) estimated masses of 
$M_1 = 1.2 \pm 0.2\, {\rm M_\odot}$ and
$M_2 = 0.99 \pm 0.03\, {\rm M_\odot}$ for the primary and secondary components,
respectively, of MU~Cen. Although still within the range observed in CVs in
general, $M_1$ is rather high; significantly higher than the average mass of 
white dwarfs in CVs which is expected to be of the order of 
$0.75\, {\rm M_\odot}$ (Knigge, 2006). However, there are two reasons 
to question the estimated masses.

First, the estimate of Friend et al.\ (1990) is based on the assumption that
the late type star is on the main sequence. However, it is well known that
the secondary stars of CVs are oversized for their mass. Therefore, using 
their Roche-lobe size together with the main sequence assumption leads to
an overestimate of their masses. The present model calculations enabled to
delimit the orbital inclination $i$ of MU~Cen. Since the radial velocity 
amplitude of the secondary, together with the orbital period and Kepler's 
third law permits to calculate the component masses as a function of $i$, it 
is now possible to constrain $M_1$ and $M_2$ without taking refuge to the main
sequence assumption. For the lower edge of the range of possible inclinations
($i=50^o$), $M_1 = 1.19 \pm 0.22 \, {\rm M_\odot}$ and 
$M_2 = 0.98 \pm 0.26 \, {\rm M_\odot}$. Here, the errors were propagated from
the errors of the mass ratio and the radial velocity amplitude as quoted by
Friend et al.\ (1990) (the error of
the orbital period has only a negligible influence). Adopting the upper edge 
($i=65^o$), $M_1 = 0.72 \pm 0.13 \, {\rm M_\odot}$ and 
$M_2 = 0.59 \pm 0.16 \, {\rm M_\odot}$. Remembering that the model results
indicaded a certain preference for higher values of $i$ within the permitted
range, the white dwarf mass may thus be close to the average mass found in
CVs. 

The second effect is harder to quantify. Friend et al.\ (1990) did 
not apply a $K$-correction (see, e.g., Wade \& Horne, 1988) to the radial 
velocity amplitude of the secondary star. This is necessary, if the absorption
lines are quenched due to irradiation on the side of the secondary which faces 
the primary component, resulting in a shift $f$ of the ``centre of gravity'' of
the absorption lines to the opposite side and thus an overestimate of its
radial velocity amplitude $K_2$. It will also have an effect on the shape of 
the observed absorption lines and thus on the measured rotational broadening
$V_{\rm rot} \sin i$.
Correcting $K_2$ implies in a smaller orbit and thus, for a given 
$P_{\rm orb}$, in smaller component masses. While the $K$-correction is a linear
function of $f$ (Wade \& Horne, 1988), the necessary correction for the 
rotational broadening may well be more complex. Since Friend et al.\ (1990) 
used uncorrected values for $K_2$ and $V_{\rm rot} \sin i$ to
calculate the mass ratio, $q$ will also require a corretion which, in turn,
has a bearing on the component masses. In order to illustrate the order of
magnitude of this effect, and assuming for simplicity that the correction
of $V_{\rm rot} \sin i$ can be neglected, $f=0.1$ would imply in a larger mass
ratio of $q=0.96$ and a primary star mass lower by 16\% in comparison to
the uncorrected masses. But, of course, $f$ is not known. In view of the
dominant contribution of the seconary star to the total light of MU~Cen it
is not expected to be large. Therefore, any $K$-correction will probably have
only a small effect on the mass ratio and the component masses.

\section{Summary}
\label{Summary}

I have shown that the quiescent light curve of MU~Cen is dominated by
ellipsoidal variations of the secondary star. This, together with the much 
longer baseline of the present observations as compared to previously  
published spectroscopy permitted a more precise determination of the orbital
period than was hitherto possible. A second persistent modulation in the 
light curve with a much smaller amplitude and a period of slightly more
than half the orbital period remains enigmatic. Flickering occurs on a rather
low level in MU~Cen, but only because of the diluting effect of the strong
contribution of the secondary star to the total light. The frequency
behaviour, as derived from a power spectrum and wavelet analysis, is similar
to that found in other quiescent dwarf novae. Flickering events may not
always be independent from one another, leading to effects reminiscent of
quasi-periodic oscillations. Model fits to the ellipsoidal variations 
yielded plausible values only for the temperature of the secondary star and for
the range of the orbital inclinations compatible with
the observations. Other important system parameters, in particular the mass
ratio, could not be constrained due to strong parameter correlations. The
range of permitted masses for the primary includes the average mass of 
white dwarfs in cataclysmic variables. All in all, the only feature which
may distinguish MU~Cen from a garden variety dwarf nova is the unexplained 
period $P_2$.

\section*{References}

\begin{description}
\parskip-0.5ex

\item
             Beckemper, S. 1995, Statistische Untersuchungen zur St\"arke des
             Flickering in kataklysmischen Ver\"anderlichen,
             Diploma thesis, M\"unster
\item
             Beuermann, K., Baraffe, I., 
             Kolb, U., \& Weichold, M. 1998, A\&A, 339, 518
\item
             Bruch, A. 1992, A\&A, 266, 237
\item Bruch, A. 1993, 
             MIRA: A Reference Guide (Astron.\ Inst.\ Univ.\ M\"unster
\item Bruch, A. 1999, AJ, 117, 3031
\item
             Bruch, A. 2003, A\&A, 409, 647
\item
             Bruch, A. 2014, A\&A, 566, A101
\item
             Bruch, A., \& Engel, A. 1994, A\&AS, 104, 79
\item
             Caceci, M.S., \& Cacheris, W.P. 
             1994, Byte, May 1984, 340
\item
             Campbell, R.D., 
             \& Shafter, A.W. 1995, ApJ, 440, 336 
\item Casares, J., 
             Mart\'{\i}nez-Pais, I.G., \& Rodr\'{\i}guez-Gil, P. 2009, 
             MNRAS, 399, 1534
\item Cenarro, A.J., 
             Peletier, R.F., 
             S\'anchez-Bl\'asquez, P., et al. 2007 MNRAS, 374, 664 
\item Chui, C.K. 1992, 
             {\it An Introduction on Wavelets}, Academic Press, San Diego 
\item
             Claret, A. 2004, A\&A, 428, 1001 
\item
             Claret, A., Bloemen, S. 
             2011, A\&A, 529, A75 
\item
             Costero, R., Echevarria, J., 
             Michel R., \& Zharikov, S. 2004, BAAS, 36, 1371 
\item
             Dunford, A., Watson, C.A., 
             Smith, R.C. 2012, MNRAS, 422, 3444
\item
             Echevarr\'{\i}a, J. 1988, 
             ApSS, 130, 103
\item
             Friend, M.T., Martin, J.S., 
             Smith, R.C., \& Jones, D.H.P. 1990, MNRAS, 246, 654,
\item
             Fritz, T., \& Bruch, A., 1998 
             A\&A, 332, 586
\item
             Horne, J.H., \& Baliunas, S.L. 
             1986, ApJ, 302, 757
\item
             Howell, S.B., Harrison, T.E., 
             Szkody, P., \& Silvestri, N.M. 2010, AJ, 139, 1771
\item
             Jawerth, B. \& Sweldens, W. 
             1994, SIAM Rev., 36, 337 
\item
             Knigge, C. 2006, MNRAS, 373, 484
\item
             Lomb, N.R. 1976, ApSS, 39, 447
\item
             Mumford, G. 1971, ApJ, 165, 369
\item
             Nohlen, C. 1995, Das Frequenzverhalten des
             Flickering von kataklysmischen Ver\"anderlichen,
             Diploma thesis, M\"unster 
\item
             Oskaki, Y., \& Kato T. 2014, 
             PASJ 66, 15
\item
             Peters, C.S., \& 
             Thorstensen, R. 2005, PASP, 117, 1386 
\item
             Rafert, J.B., Twigg, L.W. 1980, 
             MNRAS, 193, 79
\item
             Ritter, H., Kolb, U. 2003, 
             A\&A, 404, 301
\item
             Scargle, J.D. 1982, ApJ, 263, 853
\item Scargle, J.D., 
             Steiman-Cameron, T.Y., \& Young, K. 1993, ApJ, 411, L91
\item
             Schwarzenberg-Czerny, A. 1989, MNRAS, 241, 153
\item        Schwarzenberg-Czerny, A. 1991, MNRAS, 253, 189
\item
             Smith, A.J., Haswell, C.A., 
             Murray, J.R., Truss, M.R., \& Foulkes, S.B. 2007, MNRAS 378, 785
\item
             Smith, D.A., Dhillon, V.S., 
             \& Marsh, T.R. 1998, MNRAS, 296, 465
\item
             Staude, A., Schwope, A.D., 
             \& Schwarz, R. 2001, A\&A, 374, 588
\item
             Stover, R.J. 1981, ApJ, 249, 673
\item
             Thoroughgood, T.D., 
             Dhillon, V.S., Watson, C.A., et al. 2004, MNRAS, 353, 1135
\item
             Thorstensen, J.R., 
             Fenton, W.H., \& Taylor, C.J. 2004, PASP, 116, 300
\item
             Thorstensen, J.R., 
             Peters, C.S., \& Skinner, J.N. 2010, PASP, 122, 1285
\item
             Thorstensen, J.R., 
             Skinner, J.N. 2012, AJ, 144, 81
\item
             Thorstensen, J.F., 
             \& Taylor, C.J. 2001, MNRAS, 326, 1235
\item
             Vogt, N. 1976; in: Structure and Evolution of 
             close binary systems, Proc.\ IAU Symp.\ 73 (eds.: P. Eggleton et 
             al.); Reidel, Dordrecht; p.\ 147)
\item
             Vogt, N. 1981, SU UMa - Sterne und andere
             Zwergnovae, Habilitationsschrift, 
             Universit\"at Bochum
\item Vogt, N. 1983, A\&AS 53, 21
\item
            Wade, R.A., \& Horne K. 1988, 
            ApJ, 324, 411
\item
             Warner, B. 1976, in: Structure and 
             Evolution of close binary systems, Proc.\ IAU Symp.\ 73 (eds.: 
             P. Eggleton et al.); Reidel, Dordrecht, p.\ 85)
\item
             Warner, B. 2004, PASP, 116, 115
\item
             Wilson, R.E. 1979, ApJ, 234, 1054
\item
             Wilson, R.E., Devinney, E.J. 
             1971, ApJ, 166, 605
\item
             Zacharias, N., Finch, C.T., 
             Girard, T.M., et al.\ 2013, AJ, 145, 44
\item
             Zwitter, T., \& Munari, U. 
             1996, A\&AS, 117, 449

\end{description}

\end{document}